\documentclass[11pt,a4paper]{article}

\usepackage{graphicx}
\usepackage{booktabs}
\usepackage{amsmath}
\usepackage[super,sort&compress]{natbib}
\setcitestyle{citesep={,}}

\usepackage{manuscript-style}

\title{Edge intelligence for distributed acoustic \mbox{sensing} with ultracompact neural networks}
\author{
 Yuan Gao$^{1}$, Yu Tao$^{1}$, Xuzhe Zhang$^{1}$, Junyi Duan$^{1}$, Da Zhou$^{1}$, Xingcan Liu$^{1}$, Qianlin Qian$^{1}$, Shaomin Zhu$^{2}$, Xuhui Yu$^{2}$, Xinyu Fan$^{1}$, Yangyang Wan$^{1,*}$ and Zuyuan He$^{1}$\\[0.55em]
 $^{1}$State Key Laboratory of Photonics and Communications, Shanghai Jiao Tong University, Shanghai, China\\[2pt]
 $^{2}$Ningbo AllianStream Photonics Technology Co., Ltd., Ningbo, China\\[0.35em]
 $^{*}$Correspondence: Yangyang Wan\\[2pt]
 Emails: Yangyang Wan (\texttt{YangyangWan@sjtu.edu.cn})
}
\date{}

\begin{document}

\maketitle
\begin{abstract}

Distributed acoustic sensing (DAS) transforms optical fibres into dense arrays of vibration sensors for continuous, long-range monitoring with high spatial resolution. Yet converting these large-scale measurements into actionable information with compact models that retain task performance remains challenging. Existing artificial-intelligence approaches are typically optimized for specific sensing environments and tasks, but their computational requirements can hinder deployment in compact, power-constrained field systems. Here, we introduce DAS-NanoNet, a framework that automatically constructs lightweight neural networks for different DAS scenarios and tasks, together with OmniDAS, an open benchmark spanning six representative environments. Using lightweight temporal operators tailored to DAS signals, an adaptive neural architecture search strategy derives a compact network for each scenario–task pairing. Across classification, localization, compression, and denoising tasks in diverse DAS applications, DAS-NanoNet consistently yields highly compact architectures. On a six-class dataset, it achieves 99.68\% accuracy with only 1,391 trainable parameters, while delivering comparable or better performance with up to 18,098-fold fewer parameters than previously reported high-accuracy networks. Its compact architecture further enables edge deployment, achieving sub 9 $\mu$s inference latency on the deployed hardware platform across different DAS applications. DAS-NanoNet provides a route towards edge intelligence for compact, low-power and real-time DAS systems through ultracompact networks tailored to individual sensing scenarios and tasks.

\end{abstract}

\section*{Introduction}

Distributed acoustic sensing (DAS) has emerged as a powerful distributed sensing technology by exploiting optical fibres as continuous sensing media to retrieve vibration information along their entire lengths\cite{he2021review,jousset2018dynamicstrain}. 
By providing metre-scale spatial resolution over sensing distances spanning tens to hundreds of kilometres, DAS enables high-density and large-scale monitoring that is difficult to achieve with conventional sensing approaches\cite{liu2026bridgehealth}. 
These unique capabilities have stimulated growing interest in a wide range of applications, including geophysical exploration\cite{li2023asperities,jousset2022volcanic}, environmental monitoring\cite{graff2025calving,lin2024oceancurrents,walter2020glacier}, urban sensing\cite{liu2025urbansensing,cohen2025traffic}, and infrastructure health monitoring\cite{liu2026bridgehealth,tan2025aqueduct}. 
Beyond its ability to continuously capture distributed vibration signals, DAS offers a foundation for pervasive monitoring networks\cite{lindsey2019seafloor,tang2026intrusion}. Realizing these applications at the sensing edge requires accurate signal interpretation within limited computational and memory resources.

While DAS provides unprecedented access to distributed vibration information, extracting meaningful and application-specific knowledge from massive sensing data remains a critical challenge.
Signal processing plays a central role in transforming DAS from a system that merely records distributed vibrations into an intelligent sensing platform capable of detecting\cite{hernandez2022dasearthquake}, interpreting, and analysing relevant events\cite{munoz2022nearfield,li2023focalmechanisms}. 
Compared with conventional signal processing approaches, artificial intelligence (AI)-based methods have demonstrated improved performance across a wide range of DAS tasks, including event detection\cite{zhu2023phasepicking,yu2024daseventnet}, signal classification\cite{sawi2026magnitude,peng2020locomotion}, noise suppression\cite{vanDenEnde2023denoising,lapins2024dasn2n}, data compression\cite{chen2025dascompression}, and fault diagnosis\cite{wan2026physicsinformed}. 
However, the effectiveness of existing DAS-AI approaches is often confined to specific sensing scenarios and predefined tasks, limiting their ability to maintain accuracy and efficiency when acquisition conditions or deployment environments change\cite{zhu2023phasepicking,zhang2024adai,wan2026physicsinformed}. 
To improve adaptability under changing conditions, several strategies have been explored, including transfer learning\cite{shi2021transfer}, semi-supervised learning\cite{zhu2023phasepicking}, self-supervised learning with DAS-MAE\cite{duan2026dasmae}, knowledge distillation\cite{luo2025realtime}, decentralized adaptation\cite{zhang2024adai}, and physics-informed data generation\cite{wan2026physicsinformed}. 
Although these approaches can reduce the requirements for extensive annotation or complete model retraining, they generally perform adaptation within a pre-existing architecture by modifying data distributions, model parameters, or learning objectives.
Different sensing conditions and processing tasks can require different temporal receptive fields and model capacities. Consequently, a manually designed network topology that performs effectively under one sensing configuration may provide a suboptimal balance between task performance and model size in another.

The practical deployment of intelligent DAS systems is further challenged by the substantial computational burden associated with large-scale, high-rate spatiotemporal data streams, which must be processed continuously and with low latency\cite{zhu2023phasepicking}. To achieve high signal-processing and analysis performance, many existing DAS-AI approaches rely on increasingly deep and large neural networks, resulting in substantial computational and memory demands\cite{zhu2023phasepicking,duan2026dasmae}. 
For example, DAS-MAE uses a self-supervised pretraining model with approximately 28~million parameters\cite{duan2026dasmae}, while MAEPD employs a backbone with approximately 85.8~million parameters for representation learning across diverse DAS applications\cite{gui2026maepd}. HOG-MAE achieves 96.86\% accuracy in DAS-based gait recognition using an encoder with approximately 86.6~million parameters\cite{shi2026hogmae}.
Such high-performance models commonly require dedicated computing hardware with considerable power consumption, such as GPUs, which can be difficult to accommodate in deployment-constrained sensing environments, including field-based geohazard monitoring\cite{zhu2023phasepicking,luo2025realtime}. These hardware requirements also limit the scalability of DAS intelligence across geographically distributed and heterogeneous sensing scenarios. To reduce inference latency, several studies have explored deploying DAS neural networks on field-programmable gate arrays (FPGAs)\cite{luo2025realtime,liu2025rapid}.
Although these approaches demonstrate the feasibility of real-time inference, they typically rely on fixed network architectures designed in advance for specific scenarios or tasks, while even lightweight implementations still involve tens of thousands of parameters and inference latencies on the order of 100~$\mu$s\cite{luo2025realtime}.
Therefore, developing compact DAS neural networks that retain high performance across diverse scenarios and tasks is critical for resource-efficient edge intelligence and remains a major bottleneck to the widespread deployment of DAS. 

The evaluation of compact DAS networks across sensing scenarios and tasks is constrained by the limited availability of high-quality, publicly accessible datasets and standardized benchmarks\cite{tomasov2025dataset}.
Building representative and reusable datasets for DAS AI development and evaluation typically requires the extraction of sufficient high-quality and accurately labelled events from massive volumes of raw recordings, a process that is both labour-intensive and time-consuming\cite{zhu2023phasepicking,wan2026physicsinformed}. 
This challenge is further amplified when collecting DAS data across diverse sensing scenarios and tasks, as it requires dedicated interrogation systems, controlled or field experiments, and detailed acquisition procedures to capture heterogeneous signal characteristics\cite{jousset2018dynamicstrain,tomasov2025dataset}. 
Although several public DAS datasets, including PubDAS\cite{spica2023pubdas} and task-specific event classification datasets\cite{cao2023eventdataset}, have been released, some resources remain difficult to access or have incomplete annotations, variable labelling quality, missing acquisition metadata, or non-standardized data partitions.
More comprehensively documented datasets, such as that of Tomasov et al.\cite{tomasov2025dataset}, still lack joint coverage of diverse application scenarios and processing tasks.
These limitations make it difficult to compare models fairly across sites and tasks or to repurpose existing data for new applications.
Consequently, the DAS AI community still lacks a comprehensive benchmark that encompasses diverse application scenarios and processing tasks, while providing detailed acquisition metadata, traceable annotations, standardized data splits, and task-specific evaluation protocols. 
Establishing such a benchmark is essential for enabling reliable performance assessment, facilitating algorithmic development, and accelerating the practical deployment of intelligent DAS systems.

\begin{figure}[p]
 \centering
 \includegraphics[width=\linewidth]{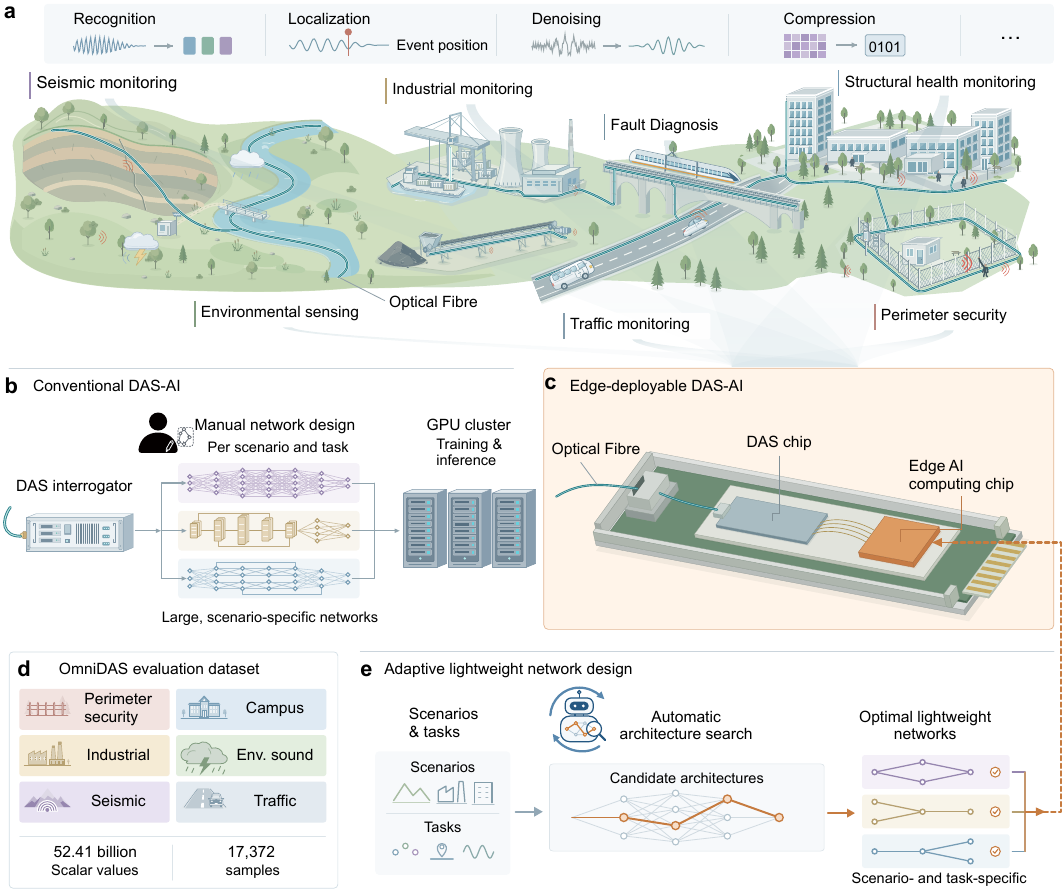}
 \caption{\textbf{From diverse DAS applications to adaptive edge intelligence.} \textbf{a}, Conceptual examples of distributed acoustic sensing applications and processing tasks, including recognition, localization, denoising and compression. \textbf{b}, Conventional processing with manually designed, scenario-specific networks and back-end GPU computing. \textbf{c}, Prospective integration of a DAS sensing chip with edge-AI computing circuitry in a compact module. \textbf{d}, The proposed open benchmark, OmniDAS, provides 17,372 native records and 52.41 billion stored scalar values from six sensing scenarios for model development and evaluation. \textbf{e}, Scenario- and task-specific architecture search selects compact networks for edge deployment.}
 \label{fig:architecture}
\end{figure}

Here, we present DAS-NanoNet, a network architecture framework designed to enable lightweight yet high-performance deep learning models for diverse DAS applications and tasks, and introduce OmniDAS, an open benchmark for AI algorithm development and standardized evaluation in the DAS research community (Fig.~\ref{fig:architecture}).
Using a search space of lightweight temporal operators tailored to DAS signals, DAS-NanoNet performs scenario- and task-adaptive architecture search to derive a highly compact network for each scenario--task pairing while maintaining competitive performance.
We demonstrate the capability of DAS-NanoNet across diverse DAS applications, including industrial structural health monitoring, perimeter security, and seismic monitoring, covering multiple tasks such as event classification, localization, data compression (including waveform reconstruction), and denoising. 
Across these scenarios and tasks, DAS-NanoNet reduces the number of network parameters by more than two to four orders of magnitude compared with conventional hand-designed neural networks, while preserving or improving task performance. 
Enabled by its extremely lightweight architecture with only thousands of parameters, DAS-NanoNet can be easily deployed on hardware circuits, achieving microsecond-scale low-latency and low-power edge signal processing. 
This capability provides a pathway towards overcoming the computational constraints of field-deployed DAS systems and enables high-speed, high-performance, multi-scenario and multi-task processing directly at the sensing edge.
In addition, OmniDAS incorporates six representative DAS application scenarios with substantially different signal characteristics and provides task-specific data partitions and evaluation methods for classification, localization, compression, and denoising tasks. 
To our knowledge, OmniDAS represents the first open DAS benchmark that covers multiple application scenarios and processing tasks, providing a common foundation for advancing data-driven DAS AI research and enabling more reliable comparisons among emerging AI approaches.

\section*{Results}

\subsection*{The OmniDAS benchmark captures heterogeneous distributed acoustic sensing}

\begin{figure}[p]
 \centering
 \includegraphics[width=\linewidth]{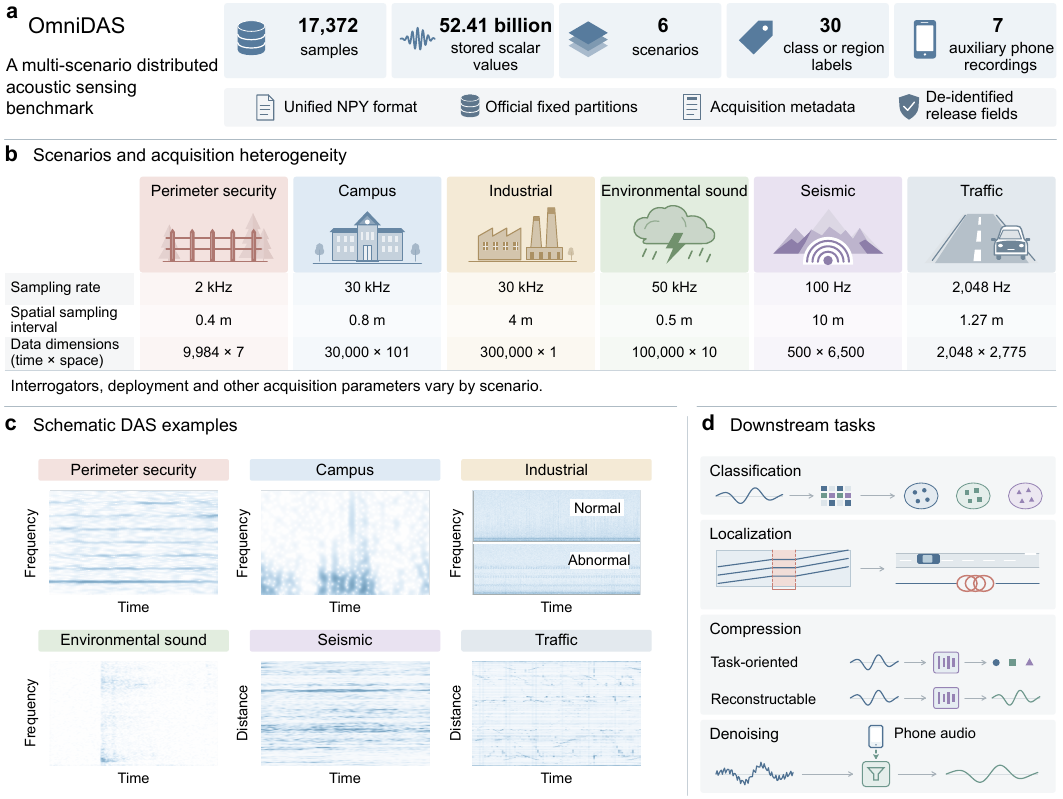}
 \caption{\textbf{OmniDAS benchmark composition, acquisition settings and tasks.} \textbf{a}, Benchmark scale: 17,372 native records, 52.41 billion stored scalar values, six scenarios and 30 class or region labels, with seven additional phone recordings as auxiliary acoustic references. Records are released in NPY format with fixed partitions, acquisition metadata and de-identified fields. \textbf{b}, Scenario-specific sampling rates, spatial sampling intervals and native data dimensions, with dimensions ordered as time $\times$ space. \textbf{c}, Schematic signal examples illustrating heterogeneous time--frequency and time--distance structures. \textbf{d}, Classification, fixed-region localization, task-oriented and waveform-reconstruction compression, and denoising. The Traffic localization target is the cable-slack region; phone audio supplies an auxiliary spectral prior for denoiser selection.}
 \label{fig:benchmark}
\end{figure}

A persistent challenge for the DAS-AI community is the lack of a suitable benchmark for systematic comparison across different algorithms. To enable a fair and consistent evaluation of DAS-NanoNet against existing methods, we introduce OmniDAS, an open benchmark for DAS intelligence. We use OmniDAS for multi-scenario and multi-task evaluations and an external six-class benchmark for additional classification comparisons. We first briefly describe the benchmark before introducing DAS-NanoNet. 
The OmniDAS benchmark contains 17,372 native records, 52.41 billion stored scalar values and 30 class or region labels (Fig.~\ref{fig:benchmark}). The six acquired scenarios are Perimeter Security, Campus, Industrial, Environmental Sound, Seismic and Traffic, and they differ in interrogator, fiber layout, temporal and spatial sampling, record shape, duration and label construction. 
In addition to straight-run fiber optic cable deployments,
OmniDAS includes recordings acquired using the mechanically robust,
highly sensitive fiber optic microphone reported by
Zhang et al.\cite{zhang2026robustmicrophone}.

Five of the six provide classification labels. In Traffic, a time-invariant binary distance label marks the cable-slack region and defines the fixed-region localization task. For the Fiber Optic Microphone deployment within the Industrial scenario, we additionally release seven phone recordings from the same deployment as an auxiliary acoustic reference for denoising. The dataset also supports a broader range of DAS signal-processing tasks, including compression and waveform reconstruction, as well as cross-scenario generalization evaluation.

Each record retains its native array shape and data type. Official fixed partitions, label provenance, acquisition metadata and de-identified release fields are specified per scenario. 

\subsection*{Design and parameter efficiency of DAS-NanoNet}
\begin{figure}[p]
 \centering
 \includegraphics[width=\linewidth]{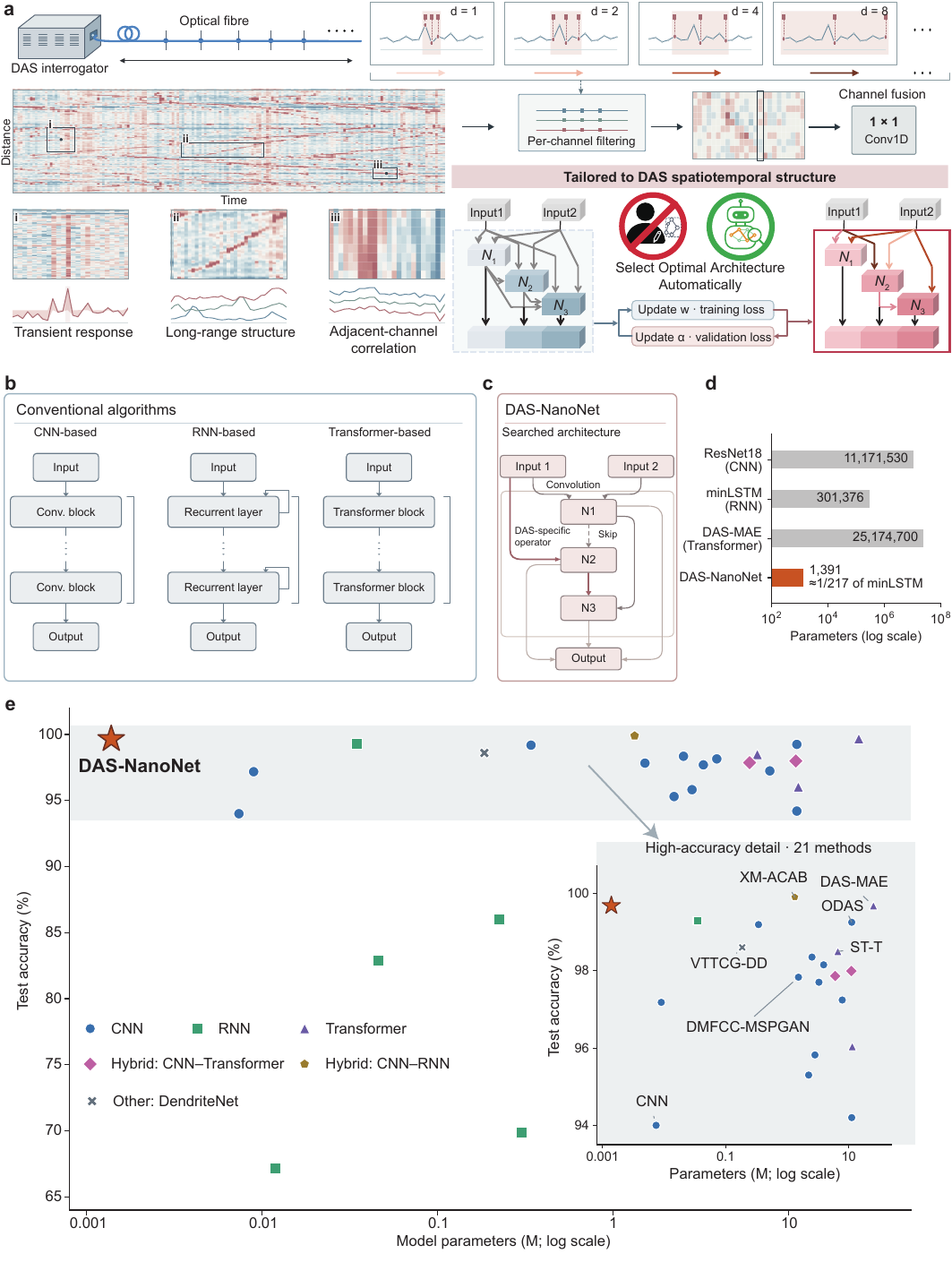}
 \caption{\textbf{DAS-NanoNet design and parameter efficiency.} \textbf{a}, DAS-informed temporal filtering with dilation rates $d$ and channel fusion guide differentiable architecture search; $w$ and $\alpha$ denote network weights and architecture parameters. \textbf{b,c}, Conventional CNN, RNN and Transformer architectures (b) and a searched DAS-NanoNet (c). \textbf{d}, Parameter counts of representative models. \textbf{e}, Reported test accuracy versus parameter count for DAS-NanoNet and 24 comparators on the external six-class dataset; the inset highlights 21 high-accuracy methods. DAS-NanoNet achieves 99.68\% accuracy with 1,391 parameters.}
 \label{fig:search-results}
\end{figure}

Unlike a manually designed network with a fixed topology, DAS-NanoNet
adaptively searches for a compact network structure suited to each sensing
scenario and task (Fig.~\ref{fig:search-results}a).
We first define a compact set of candidate operations, including operators
designed specifically for the spatiotemporal characteristics of DAS signals.
These operators filter the signal within each sensing channel over different
time spans and then combine information across channels.
This allows them to capture brief disturbances, patterns over longer time
intervals and relationships between sensing channels with few parameters.
During the search, the candidate operations are trained together within a
single network, with adjustable selection scores controlling how their
outputs are combined.
These scores are updated according to performance on the target task,
guiding the selection of operations and connections suited to the DAS
signals being processed\cite{liu2019darts,xue2022grafting}.
The highest-scoring choices are then retained to form a compact network,
which is retrained for the target application.
The search is performed independently for each sensing scenario and task,
allowing the network structure to adapt to different signal characteristics
and processing requirements.
The architecture-search formulation is described in Methods.

On the external six-class dataset introduced by Cao et al.\cite{cao2023eventdataset}, the searched architecture contained only 1,391 trainable parameters and reached 99.68\% test accuracy (Fig.~\ref{fig:search-results}). Among the 24 comparison models, DAS-NanoNet outperformed 22 in reported accuracy and matched DAS-MAE while using approximately 18,098-fold fewer parameters\cite{duan2026dasmae}. Only XM-ACAB achieved a higher accuracy, by 0.22 percentage points, but required approximately 957-fold more parameters\cite{li2024deeplearning}. These results place the searched network in the strongest accuracy--parameter operating region of the comparison and show that architecture search can surpass most fixed designs while retaining an exceptionally small model footprint. 

\subsection*{DAS-NanoNet enables lightweight processing across sensing scenarios and tasks}

\begin{figure}[p]
 \centering
 \includegraphics[width=\linewidth]{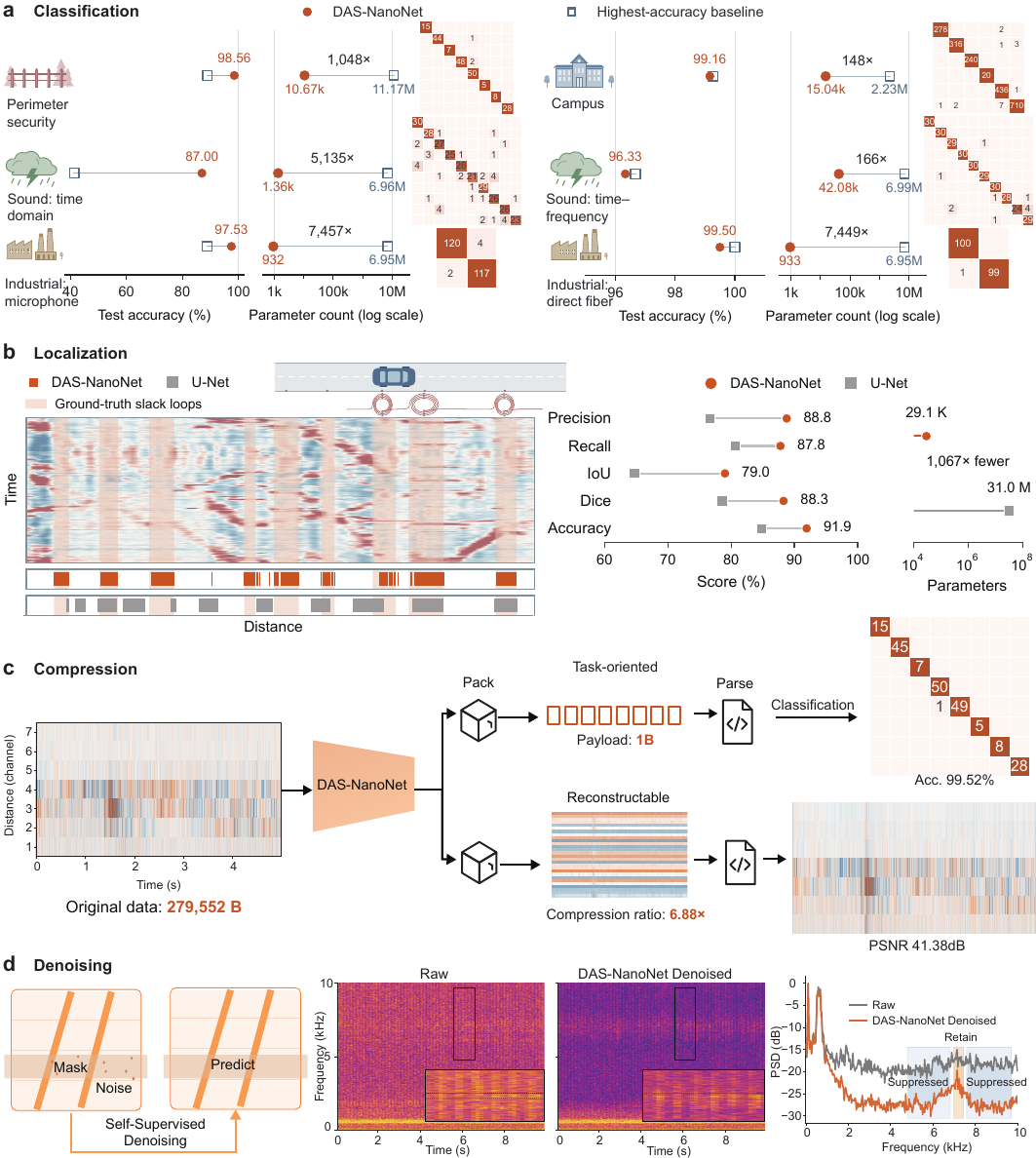}
 \caption{\textbf{DAS-NanoNet across classification, localization, compression and denoising.} \textbf{a}, Classification accuracy, parameter counts and DAS-NanoNet confusion matrices across six conditions. Orange circles denote DAS-NanoNet; open blue squares denote the highest-accuracy baselines. Matrix entries give record counts; rows and columns denote true and predicted classes, respectively. \textbf{b}, Traffic cable-slack localization and comparison with U-Net. Peach shading marks ground-truth slack regions; orange and grey strips show DAS-NanoNet and U-Net predictions. Plots compare localization metrics and parameter counts. \textbf{c}, Task-oriented and waveform-reconstruction compression of a 279,552-byte Perimeter Security record. Classification from the 1-byte task payload achieves 99.52\% test accuracy; the complete 25-byte stream gives 11,182.08-fold compression. The displayed waveform-reconstruction example achieves 6.88-fold compression and 41.38~dB peak signal-to-noise ratio (PSNR). \textbf{d}, Masked self-supervised denoising for the Industrial Fiber Optic Microphone deployment: raw and denoised spectrograms, enlarged patches and power spectral density (PSD). Grey and orange curves denote raw and denoised signals; yellow and blue shading mark the retained signal band and suppressed flanking bands.}
 \label{fig:cross-task}
 \label{fig:multicondition-results}
\end{figure}

Unlike previous approaches that require network architectures to be manually tailored to specific scenarios and tasks, DAS-NanoNet automatically searches for an optimal compact architecture, enabling it to adapt across diverse sensing scenarios and processing tasks.
We next evaluated the performance and model size of DAS-NanoNet across different sensing scenarios and processing tasks (Fig.~\ref{fig:cross-task}). 
For classification, we considered six conditions spanning Perimeter Security, Campus, two Industrial deployments and two Environmental Sound representations (Fig.~\ref{fig:multicondition-results}a). Comparisons included MobileNetV2\cite{sandler2018mobilenetv2}, DenseNet-121\cite{huang2017densenet} and ResNet18\cite{he2016resnet}. In every condition, the searched DAS-NanoNet either exceeded the conventional baselines or remained within 0.50 percentage points of the highest reported accuracy, while using 148--7,457-fold fewer parameters than the corresponding highest-accuracy comparator.

For Perimeter Security, DAS-NanoNet reached 98.56\% accuracy with 10,667 parameters, exceeding the strongest conventional baseline, ResNet18, by 9.62 percentage points. On Campus, the 15,035-parameter model achieved 99.16\% accuracy, within 0.10 percentage points of MobileNetV2 while using approximately 148-fold fewer parameters. Within the Industrial scenario, the Fiber Optic Microphone and Straight-run Fiber Optic Cable deployments reached 97.53\% and 99.50\% accuracy with only 932 and 933 parameters, respectively. The former exceeded DenseNet-121 by 8.64 percentage points, whereas the latter remained within 0.50 percentage points of DenseNet-121 and ResNet18.

Environmental Sound presented less distinctive class-specific patterns in the time domain. Despite this difficulty, DAS-NanoNet achieved 87.00\% accuracy and 86.88\% Macro-F1 directly from time-domain inputs with only 1,355 parameters. The corresponding accuracies of MobileNetV2, DenseNet-121 and ResNet18 were 30.33\%, 41.33\% and 31.00\%, respectively. DAS-NanoNet therefore exceeded the strongest time-domain baseline by 45.67 percentage points while using approximately 5,135-fold fewer parameters. With a short-time Fourier transform representation, an independently searched model reached 96.33\% accuracy with 42,076 parameters, within 0.34 percentage points of DenseNet-121 while using approximately 166-fold fewer parameters. These results show that scenario-specific search retained a compact model footprint across heterogeneous deployments and input representations, including signals whose class-specific temporal patterns were less apparent. 

Beyond classification, the same search strategy selected separate compact networks for localization, compression and denoising (Fig.~\ref{fig:cross-task}b--d). For Traffic cable-slack localization, the 29,086-parameter DAS-NanoNet achieved 79.03\% IoU, compared with 64.73\% for a U-Net\cite{ronneberger2015unet} containing approximately 1,067 times more parameters (Fig.~\ref{fig:cross-task}b).

In Perimeter Security, the 10,839-parameter task-oriented compression model encoded each 279,552-byte input record into a 1-byte task payload comprising four 2-bit codes. Together with a 20-byte header and a 4-byte CRC32 checksum, the complete stream occupied 25 bytes, corresponding to an 11,182.08-fold reduction in record size. The task-code confusion matrix in Fig.~\ref{fig:cross-task}c shows that classification from the compressed code achieved 99.52\% accuracy and 99.75\% Macro-F1 on 208 test records. For waveform-reconstruction compression, the 13,419-parameter model achieved a 3.74-fold reduction with 28.556~dB mean PSNR and 0.858 mean Pearson correlation across the same 208 test records. The reconstructed waveforms also supported downstream recognition, with the frozen classifier achieving 71.63\% accuracy and 68.44\% Macro-F1 without retraining on reconstructed inputs. Together with the reconstruction metrics, these results show that the compact model reduced data volume while retaining waveform structure and information useful for classification.

For self-supervised denoising in the Industrial scenario's Fiber Optic Microphone deployment, the selected 10,561-parameter network reduced median signal RMS by 1.028~dB for abnormal records and 0.869~dB for normal records, while retaining a median raw--denoised correlation of 0.925. In Fig.~\ref{fig:cross-task}d, the paired spectrograms and enlarged patches show reduced background spectral energy while the dominant anomaly-related time--frequency structure remains visible. The corresponding power spectra show 2.66~dB greater attenuation in the flanking background bands than in the anomaly band. These time--frequency and spectral comparisons support suppression of background noise while preserving the principal anomaly structure. On the same 243 Industrial test records, the frozen anomaly classifier achieved 77.37\% accuracy on the denoised signals without retraining on denoised inputs. 

\subsection*{Lightweight DAS-NanoNet enables edge deployment}

\begin{figure}[p]
 \centering
 \includegraphics[width=\linewidth]{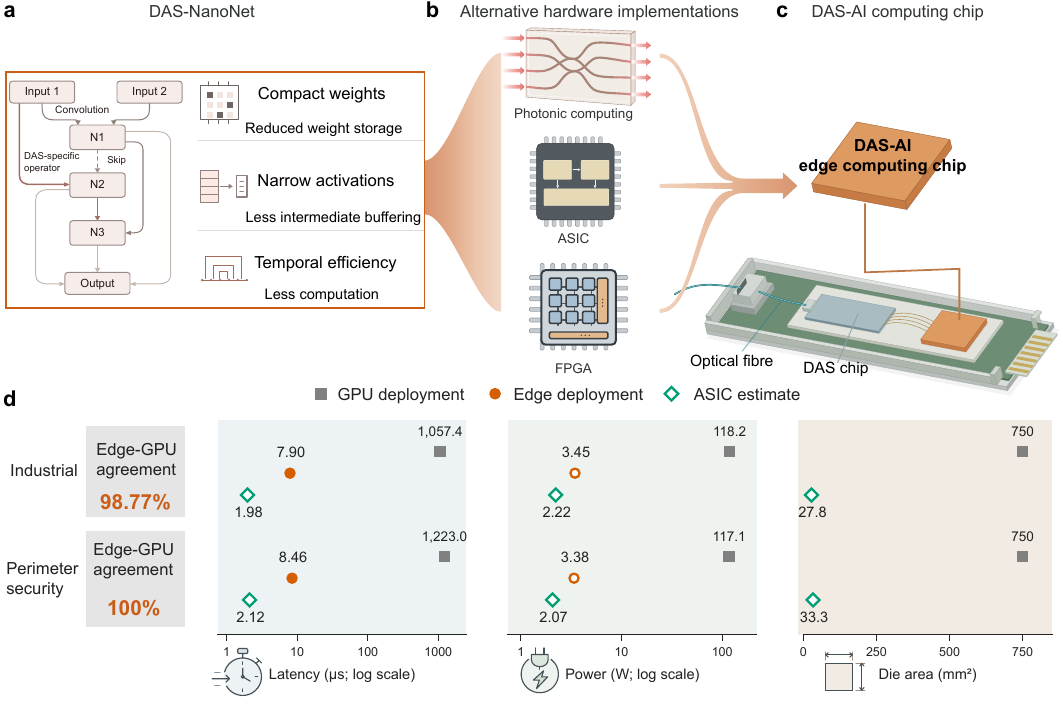}
 \caption{\textbf{Compact DAS-NanoNet architectures enable edge processing.} \textbf{a}, Compact weights, narrow activations and temporal processing reduce storage, buffering and computation requirements. \textbf{b,c}, Candidate hardware platforms and a prospective integrated DAS-AI module. FPGA deployment provides the experimental proof of concept; ASIC and photonic implementations are future possibilities. \textbf{d}, Batch-one, 16-point-input deployment results for Industrial Fiber Optic Microphone and Perimeter Security. FPGA core latencies are 7.90 and 8.46~$\mu$s; GPU forward times are 1,057.4 and 1,223.0~$\mu$s. Edge--GPU prediction agreement is 98.77\% (240/243) and 100\% (208/208). Latency, power and die-area comparisons include FPGA implementations, GPU references and ASIC projections.}
 \label{fig:fpga}
\end{figure}

The ultracompact architecture of DAS-NanoNet reduces the requirements
for weight storage, intermediate buffering and computation, thereby
lowering the barrier to edge deployment and enabling a broader range
of hardware implementations.
For example, a silicon photonic processor has been experimentally
evaluated using a neural network with 1,350 trainable
parameters\cite{roumpos2024silicon}, comparable in parameter count
to our DAS-NanoNet.
This motivates exploring photonic implementations of DAS-NanoNet
with appropriate adaptation to the hardware.
Its small parameter count also reduces the weight-storage burden,
making DAS-NanoNet a promising candidate for application-specific
integrated circuit (ASIC) implementation, where memory access
and data movement are major contributors to energy
consumption\cite{chen2017eyeriss}.
Silicon photonic integrated DAS interrogators have also been demonstrated \cite{jin2024silicon}, and their future hybrid integration with dedicated edge-deployed DAS-AI chips could enable a new generation of compact and modular DAS systems (Fig.~\ref{fig:fpga}a--c).

Constrained by available experimental resources and to provide a practical hardware demonstration, we deploy DAS-NanoNet on an FPGA and evaluate the resulting edge implementation in two distinct application scenarios.
Both evaluated deployments used 16-point inputs at batch size one. The models for the Industrial scenario's Fiber Optic Microphone deployment and for Perimeter Security achieved FPGA core inference latencies of 7.90 and 8.46~$\mu$s, respectively, compared with GPU forward latencies of 1.057 and 1.223~ms. These values correspond to approximately 134- and 145-fold reductions in model-execution latency, demonstrating microsecond-scale inference for both sensing scenarios (Fig.~\ref{fig:fpga}d).

FPGA and GPU predictions agreed for 240 of 243 Industrial records (98.77\%) and all 208 Perimeter Security records (100\%). 
The estimated FPGA on-chip powers were 3.454 and 3.381~W, respectively. With an estimated on-chip power budget of approximately 3.4~W and core latencies below 9~$\mu$s, both implementations support low-power, low-latency DAS inference at the edge. 

We anticipate that DAS-NanoNet could be deployed on ASIC and photonic computing platforms in the future, enabling further reductions in inference latency and power consumption.
Resource-based ASIC projections further explore the potential for dedicated-chip integration, with latency, power and die-area estimates summarized in Fig.~\ref{fig:fpga}d. 

\section*{Conclusion}

In conclusion, we have developed DAS-NanoNet, a framework that uses scenario- and task-adaptive architecture search to construct ultracompact neural networks for edge intelligence in DAS. To support the development of such lightweight, adaptive models and enable consistent and fair comparisons with existing DAS-AI approaches, we further establish and openly release OmniDAS, a standardized benchmark spanning six heterogeneous DAS application scenarios and multiple processing tasks. OmniDAS comprises 17,372 native records, 52.41 billion stored values and 30 labels, with dedicated tracks for classification, localization, compression and denoising. By providing diverse DAS data, task-specific data partitions and standardized evaluation settings within a common benchmark, OmniDAS offers a shared resource for developing, evaluating and comparing AI algorithms across sensing scenarios and tasks, thereby supporting more reproducible and systematic progress in the DAS-AI community.

By searching over lightweight temporal operators tailored to DAS signals, DAS-NanoNet independently derives compact networks for different sensing scenarios and processing tasks. On the external six-class classification benchmark, a network with only 1,391 trainable parameters achieved 99.68\% accuracy, matching DAS-MAE with 18,098-fold fewer parameters. Across six OmniDAS classification settings, independently searched architectures reduced parameter counts by 148--7,457 fold relative to the highest-accuracy conventional comparators while retaining competitive reported accuracy. The same framework extends beyond classification to localization, compression and denoising: cable-slack localization improved IoU by 14.30 percentage points over U-Net with an approximately 1,067-fold smaller model; a 25-byte Perimeter Security stream achieved an 11,182.08-fold compression ratio while retaining 99.75\% Macro-F1; waveform compression achieved a mean PSNR of 28.556~dB at a 3.74-fold compression ratio; and self-supervised denoising was implemented with only 10,561 parameters. The resulting lightweight architectures are also readily amenable to edge deployment, with FPGA implementations for the Industrial Fiber Optic Microphone and Perimeter Security scenarios achieving core inference latencies of 7.90 and 8.46~$\mu$s, respectively.

The combination of competitive task performance, scenario- and task-adaptive architecture design, and a small parameter footprint provides a route towards resource-constrained edge intelligence. Beyond reducing the dependence of field-deployed DAS systems on bulky computing hardware, such lightweight intelligence may facilitate the on-chip integration of signal demodulation and processing functions. Further integration of electronic demodulation and edge-AI circuitry with emerging optical detection chips could provide a basis for heterogeneous optoelectronic DAS architectures in which optical sensing and real-time intelligent processing are implemented within compact, low-power platforms. Such developments may advance next-generation DAS towards broadly deployable edge-intelligent sensing systems for applications including structural health monitoring, perimeter security, seismic observation and geological hazard monitoring.

\section*{Methods}

\subsection*{DAS-NanoNet architecture search}

DAS-NanoNet independently searches the network topology for each scenario--task pairing. The input representation and task head are fixed before each search, while the operations and connections in the intermediate temporal layers are selected through task-specific architecture search. For a candidate edge from layer $i$ to layer $j$, the continuous supernetwork output is
\begin{equation}
\bar{o}^{(i,j)}(x_i)=\sum_{o\in\mathcal{O}}\frac{\exp(\alpha_o^{(i,j)})}{\sum_{o'\in\mathcal{O}}\exp(\alpha_{o'}^{(i,j)})}\,o(x_i),
\label{eq:main-darts-mixture}
\end{equation}
where $\mathcal{O}$ is the task-specific operator set and $\alpha$ contains the architecture parameters. Following differentiable architecture search\cite{liu2019darts}, we optimize the architecture parameters $\alpha$ using the validation loss and the network weights $w$ using the training loss. The bi-level objective is
\begin{equation}
\min_{\alpha}\;\mathcal{L}_{\mathrm{val}}\!\left(w^*(\alpha),\alpha\right),\qquad
w^*(\alpha)=\arg\min_w\mathcal{L}_{\mathrm{train}}(w,\alpha).
\label{eq:main-darts-bilevel}
\end{equation}
The highest-scoring non-null operators and connections are retained to form a network with a fixed topology, which is retrained and frozen before test evaluation. Classification, localization, compression and denoising use independent searches and validation metrics.

\subsection*{Traffic cable-slack fixed-region localization}

The cable-slack target comprised 28 fixed physical distance intervals along one monitored traffic route. Each raw record was converted to a low-frequency distance representation by temporal centering, distance-wise RMS normalization and averaging Fourier magnitudes over $1\leq f<20$~Hz; the resulting waterfall was tiled and processed by a searched temporal encoder. A distance head produced one binary logit per spatial position, $\hat{y}_d=\sigma(h_\theta(z)_d)$. For binary decisions, performance was summarized using
\begin{align}
\mathrm{Precision}&=\frac{\mathrm{TP}}{\mathrm{TP}+\mathrm{FP}}, &
\mathrm{Recall}&=\frac{\mathrm{TP}}{\mathrm{TP}+\mathrm{FN}},\\
\mathrm{IoU}&=\frac{\mathrm{TP}}{\mathrm{TP}+\mathrm{FP}+\mathrm{FN}}, &
\mathrm{Dice}&=\frac{2\mathrm{TP}}{2\mathrm{TP}+\mathrm{FP}+\mathrm{FN}}.
\end{align}
The searched model was compared with U-Net\cite{ronneberger2015unet}. Reported road intersections were contextual metadata and were not used as labels or model inputs. 

\subsection*{Perimeter Security compression}

Task-oriented compression projected the classifier's pooled representation $h$ to a compact latent code, $z=\tanh(W h+b)$, quantized the code to signed $b$-bit values with a straight-through estimator\cite{bengio2013estimating}, and classified the dequantized code $\hat{y}=c_\phi(\hat{z})$. The operating point was selected on the validation split as the smallest serialized stream among configurations within 0.01 Macro-F1 of the validation best. The selected DAS-NanoNet-4D-2bit-TC serialized four 2-bit codes with a 20-byte header and 4-byte CRC32.

Waveform-reconstruction compression encoded and quantized a latent representation before decoding the native waveform shape\cite{theis2017lossy}. Its training objective was
\begin{equation}
\mathcal{L}_{\mathrm{WR}}=\mathrm{MSE}(\hat{x},x)+0.1\,\mathrm{MSE}(\Delta\hat{x},\Delta x),
\label{eq:main-waveform-loss}
\end{equation}
where $\Delta x_{c,t}=x_{c,t+1}-x_{c,t}$ and $\Delta\hat{x}_{c,t}=\hat{x}_{c,t+1}-\hat{x}_{c,t}$. Reconstruction quality was evaluated with
\begin{equation}
\mathrm{PSNR}=10\log_{10}\frac{(\max(x)-\min(x))^2}{\mathrm{MSE}},\qquad
\mathrm{NMSE}=\frac{\sum_i(\hat{x}_i-x_i)^2}{\sum_i x_i^2}.
\label{eq:main-compression-metrics}
\end{equation}
The formal waveform model used 28 latent channels and 5-bit quantization. 

\subsection*{Industrial self-supervised denoising}

Building on self-supervised DAS denoising with jDAS\cite{vanDenEnde2023denoising}, we used masked self-reconstruction for the Industrial scenario's Fiber Optic Microphone deployment. The single-channel temporal masking scheme was inspired by Noise2Void\cite{krull2019noise2void}. A binary mask $m_t$ was sampled with probability 0.1; hidden samples were replaced by the mean of their left and right neighbours, and the network received the corrupted waveform and visibility channel $1-m$. The masked loss was
\begin{equation}
\mathcal{L}_{\mathrm{mask}}(\theta)=
\frac{\sum_t m_t\left[f_\theta(\tilde{x},1-m)_t-x_t\right]^2}{\sum_t m_t}.
\label{eq:main-denoising-mask}
\end{equation}
Candidate denoisers were first ranked by validation masked MSE. A weak phone-guided spectral prior was then used to rerank the five best candidates without supplying clean targets or noise labels:
\begin{equation}
s_{\mathrm{select}}=-\mathcal{L}_{\mathrm{mask}}+0.05s_{\mathrm{phone}},
\label{eq:main-denoising-selection}
\end{equation}
where the largest score was selected. The selected model was retrained and frozen before test evaluation. 

\subsection*{Edge deployment}

Independent models for the Industrial scenario's Fiber Optic Microphone deployment and for Perimeter Security were implemented on a Xilinx VCU118 Virtex UltraScale+ evaluation board using high-level synthesis\cite{duarte2018hls4ml,aarrestad2021hls4ml}. Each native sequence was cropped to 16 points along its sequence axis before HLS conversion and board inference. Both deployments used a 50-MHz clock, reuse factor 1, AXI-Lite control and an 18-bit fixed-point format with 8 integer bits. Physical-board and HLS comparisons used the same corresponding fixed-point inputs. Fixed-point selection was based on a quantization sweep.

\section*{Data Availability}

The OmniDAS dataset will be made publicly available on Zenodo upon publication of this article. The DOI \href{https://doi.org/10.5281/zenodo.21992120}{10.5281/zenodo.21992120} has been reserved for the dataset. A benchmark overview and scenario descriptions are available at \url{https://gaoyuan-mr.github.io/OmniDAS/}.

\section*{Code Availability}

The code used in this study is available from the corresponding author upon reasonable request.

\section*{Acknowledgements}
This work is financially supported by the National Natural Science Foundation of China (NSFC) under Grant No. 62405178, 62435004.

\section*{Author Contributions}
Y.G. and Y.W. conceived and designed the study. Y.G. developed the methods, performed the experiments and data analysis, and wrote the manuscript. Y.T., X.Z., J.D., D.Z., X.L., S.Z. and X.Y. contributed data. Q.Q. supported the edge deployment. X.F., Y.W. and Z.H. supervised the research and reviewed and edited the manuscript.

\section*{Competing Interests}
The authors declare no competing interests.

\bibliographystyle{naturemag-manuscript}
\bibliography{references}

@article{he2021review,
  author  = {He, Zuyuan and Liu, Qingwen},
  title   = {Optical fiber distributed acoustic sensors: A review},
  journal = {Journal of Lightwave Technology},
  volume  = {39},
  pages   = {3671--3686},
  year    = {2021},
  doi     = {10.1109/JLT.2021.3059771}
}

@article{li2023asperities,
  author  = {Li, Jiaxuan and Kim, Taeho and Lapusta, Nadia and Biondi, Ettore and Zhan, Zhongwen},
  title   = {The break of earthquake asperities imaged by distributed acoustic sensing},
  journal = {Nature},
  volume  = {620},
  number  = {7975},
  pages   = {800--806},
  year    = {2023},
  doi     = {10.1038/s41586-023-06227-w}
}

@article{lindsey2019seafloor,
  author  = {Lindsey, Nathaniel J. and Dawe, T. Craig and Ajo-Franklin, Jonathan B.},
  title   = {Illuminating seafloor faults and ocean dynamics with dark fiber distributed acoustic sensing},
  journal = {Science},
  volume  = {366},
  number  = {6469},
  pages   = {1103--1107},
  year    = {2019},
  doi     = {10.1126/science.aay5881}
}

@article{graff2025calving,
  author  = {Gr{\"a}ff, Dominik and Lipovsky, Bradley Paul and Vieli, Andreas and Dachauer, Armin and Jackson, Rebecca and Farinotti, Daniel and Schmale, Julia and Ampuero, Jean-Paul and Berg, Eric and Dannowski, Anke and Kneib-Walter, Andrea and K{\"o}pfli, Manuela and Kopp, Heidrun and van der Loo, Enrico and {Mata Flores}, Daniel and Mercerat, Diego and Moser, Raphael and Sladen, Anthony and Walter, Fabian and Wasser, Diego and Welty, Ethan and Wetter, Selina and Williams, Ethan F.},
  title   = {Calving-driven fjord dynamics resolved by seafloor fibre sensing},
  journal = {Nature},
  volume  = {644},
  number  = {8076},
  pages   = {404--412},
  year    = {2025},
  doi     = {10.1038/s41586-025-09347-7}
}

@article{chen2025dascompression,
  author  = {Chen, Yangkang and Saad, Omar M. and Chen, Yunfeng and Savvaidis, Alexandros},
  title   = {Deep learning for seismic data compression in distributed acoustic sensing},
  journal = {IEEE Transactions on Geoscience and Remote Sensing},
  volume  = {63},
  pages   = {1--14},
  year    = {2025},
  doi     = {10.1109/TGRS.2025.3526933}
}

@article{jousset2018dynamicstrain,
  author  = {Jousset, Philippe and Reinsch, Thomas and Ryberg, Trond and Blanck, Hanna and Clarke, Andy and Aghayev, Rufat and Hersir, Gylfi P. and Henninges, Jan and Weber, Michael and Krawczyk, Charlotte M.},
  title   = {Dynamic strain determination using fibre-optic cables allows imaging of seismological and structural features},
  journal = {Nature Communications},
  volume  = {9},
  pages   = {2509},
  year    = {2018},
  doi     = {10.1038/s41467-018-04860-y}
}

@article{liu2025urbansensing,
  author  = {Liu, Jingxiao and Li, Haipeng and Noh, Hae Young and Santi, Paolo and Biondi, Biondo and Ratti, Carlo},
  title   = {Urban sensing using existing fiber-optic networks},
  journal = {Nature Communications},
  volume  = {16},
  pages   = {3091},
  year    = {2025},
  doi     = {10.1038/s41467-025-57997-y}
}

@article{liu2026bridgehealth,
  author  = {Liu, Jingxiao and Aggarwal, Jatin and Hwang, Doyun and Lee, Jigu and Wu, Yuyan and Yin, Fu and Li, Haipeng and Santi, Paolo and Sohn, Hoon and Biondi, Biondo and Ratti, Carlo and Noh, Hae Young},
  title   = {Bridge health monitoring using existing telecommunication fiber-optic networks},
  journal = {Nature Communications},
  volume  = {17},
  number  = {1},
  pages   = {9966},
  year    = {2026},
  doi     = {10.1038/s41467-026-76935-0}
}

@article{cohen2025traffic,
  author  = {Cohen, Khen and Hen, Liav and Lellouch, Ariel},
  title   = {A fiber-optic traffic monitoring network trained with video inputs},
  journal = {Scientific Reports},
  volume  = {15},
  pages   = {28954},
  year    = {2025},
  doi     = {10.1038/s41598-025-14928-7}
}

@article{spica2023pubdas,
  author  = {Spica, Zack J. and others},
  title   = {{PubDAS}: A public distributed acoustic sensing datasets repository for geosciences},
  journal = {Seismological Research Letters},
  volume  = {94},
  pages   = {983--998},
  year    = {2023},
  doi     = {10.1785/0220220279}
}

@article{tomasov2025dataset,
  author  = {Tomasov, Adrian and Zaviska, Pavel and Dejdar, Petr and Klicnik, Ondrej and Horvath, Tomas and Munster, Petr},
  title   = {Comprehensive dataset for event classification using distributed acoustic sensing ({DAS}) systems},
  journal = {Scientific Data},
  volume  = {12},
  pages   = {793},
  year    = {2025},
  doi     = {10.1038/s41597-025-05088-4}
}

@article{zhu2023phasepicking,
  author  = {Zhu, Weiqiang and Biondi, Ettore and Li, Jiaxuan and Yin, Jiuxun and Ross, Zachary E. and Zhan, Zhongwen},
  title   = {Seismic arrival-time picking on distributed acoustic sensing data using semi-supervised learning},
  journal = {Nature Communications},
  volume  = {14},
  pages   = {8192},
  year    = {2023},
  doi     = {10.1038/s41467-023-43355-3}
}

@article{vanDenEnde2023denoising,
  author  = {van den Ende, Martijn and Lior, Itzhak and Ampuero, Jean-Paul and Sladen, Anthony and Ferrari, Andr\'e and Richard, C\'edric},
  title   = {A self-supervised deep learning approach for blind denoising and waveform coherence enhancement in distributed acoustic sensing data},
  journal = {IEEE Transactions on Neural Networks and Learning Systems},
  volume  = {34},
  number  = {7},
  pages   = {3371--3384},
  year    = {2023},
  doi     = {10.1109/TNNLS.2021.3132832}
}

@inproceedings{liu2019darts,
  author    = {Liu, Hanxiao and Simonyan, Karen and Yang, Yiming},
  title     = {{DARTS}: Differentiable architecture search},
  booktitle = {International Conference on Learning Representations},
  year      = {2019},
  url       = {https://openreview.net/forum?id=S1eYHoC5FX}
}

@article{xue2022grafting,
  author  = {Xue, Xizhe and Zhang, Haokui and Fang, Bei and Bai, Zongwen and Li, Ying},
  title   = {Grafting Transformer on Automatically Designed Convolutional Neural Network for Hyperspectral Image Classification},
  journal = {IEEE Transactions on Geoscience and Remote Sensing},
  volume  = {60},
  pages   = {1--16},
  year    = {2022},
  doi     = {10.1109/TGRS.2022.3180685}
}

@article{luo2025realtime,
  author  = {Luo, Zhongyao and Wu, Hao and Ge, Zhao and Tang, Ming},
  title   = {Real-Time Event Recognition of Long-Distance Distributed Vibration Sensing With Knowledge Distillation and Hardware Acceleration},
  journal = {IEEE Internet of Things Journal},
  volume  = {12},
  number  = {11},
  pages   = {17896--17905},
  year    = {2025},
  doi     = {10.1109/JIOT.2025.3539339}
}

@article{liu2025rapid,
  title={Rapid edge-computing for intelligent fiber-optic {DAS}},
  author={Liu, Yiyu and Wu, Yongxin and Jiao, Xiben and Li, Xingbin and Wu, Huijuan and Zhou, Jun and Xu, Zhen and Yu, Xiao},
  journal={IEEE Sensors Journal},
  volume={25},
  number={10},
  pages={17062--17071},
  year={2025},
  publisher={IEEE},
  doi={10.1109/JSEN.2025.3554221}
}

@article{duarte2018hls4ml,
  author  = {Duarte, Javier and others},
  title   = {Fast inference of deep neural networks in {FPGAs} for particle physics},
  journal = {Journal of Instrumentation},
  volume  = {13},
  number  = {07},
  pages   = {P07027},
  year    = {2018},
  doi     = {10.1088/1748-0221/13/07/P07027}
}

@article{aarrestad2021hls4ml,
  author  = {Aarrestad, Thea and others},
  title   = {Fast convolutional neural networks on {FPGAs} with hls4ml},
  journal = {Machine Learning: Science and Technology},
  volume  = {2},
  number  = {4},
  pages   = {045015},
  year    = {2021},
  doi     = {10.1088/2632-2153/ac0ea1}
}

@article{cao2023eventdataset,
  author  = {Cao, Xiaomin and Su, Yunsheng and Jin, Zhiyan and Yu, Kuanglu},
  title   = {An open dataset of {$\phi$-OTDR} events with two classification models as baselines},
  journal = {Results in Optics},
  volume  = {10},
  pages   = {100372},
  year    = {2023},
  doi     = {10.1016/j.rio.2023.100372}
}

@article{duan2026dasmae,
  author  = {Duan, Junyi and Chen, Jiageng and He, Zuyuan},
  title   = {{DAS-MAE}: A Self-Supervised Framework for Universal and High-Performance Representation Learning of Distributed Acoustic Sensing},
  journal = {Journal of Lightwave Technology},
  volume  = {44},
  number  = {7},
  pages   = {2544--2556},
  year    = {2026},
  doi     = {10.1109/JLT.2025.3630138}
}

@article{gui2026maepd,
  author  = {Gui, Kun and Ren, Hongliang and Shi, Shang and Lu, Jin and Yu, Changqiu and Cao, Quanjun and Gu, Guomin and Xuan, Qi},
  title   = {{MAEPD}: A Foundation Model for Distributed Acoustic Sensing Signal Recognition via Masked Autoencoder Pre-Training and Adapter-Based Prompt Tuning},
  journal = {Sensors},
  volume  = {26},
  number  = {7},
  pages   = {2057},
  year    = {2026},
  doi     = {10.3390/s26072057}
}

@article{shi2026hogmae,
  author  = {Shi, Shang and Gui, Kun and Wang, Pengcheng and Ren, Hongliang and Lu, Jin and Yu, Changqiu and Cao, Quanjun and Gu, Guomin and Xuan, Qi},
  title   = {Gait-Based Person Identification in the {DAS} System Using {HOG-MAE}},
  journal = {IEEE Access},
  volume  = {14},
  pages   = {58282--58303},
  year    = {2026},
  doi     = {10.1109/ACCESS.2026.3683725}
}

@article{li2024deeplearning,
  author  = {Li, Yujiao and Hu, Liqin and Yu, Kuanglu},
  title   = {Unleashing potentials with deep learning: decoding the complex events for distributed fiber optic sensing applications},
  journal = {Science China Information Sciences},
  volume  = {67},
  number  = {5},
  pages   = {159402},
  year    = {2024},
  doi     = {10.1007/s11432-023-3985-y}
}

@article{wan2026physicsinformed,
  author    = {Wan, Yangyang and Wang, Haotian and Yu, Xuhui and Chen, Jiageng and Fan, Xinyu and He, Zuyuan},
  title     = {Towards a physics-informed network paradigm with data generation and background noise removal for different distributed acoustic sensing applications},
  journal   = {Light: Science \& Applications},
  volume    = {15},
  number    = {1},
  pages     = {281},
  year      = {2026},
  publisher = {Springer Science and Business Media LLC},
  doi       = {10.1038/s41377-026-02295-5}
}

@article{peng2020locomotion,
  author  = {Peng, Zhaoqiang and Wen, Hongqiao and Jian, Jianan and Gribok, Andrei and Wang, Mohan and Huang, Sheng and Liu, Hu and Mao, Zhi-Hong and Chen, Kevin P.},
  title   = {Identifications and classifications of human locomotion using {Rayleigh}-enhanced distributed fiber acoustic sensors with deep neural networks},
  journal = {Scientific Reports},
  volume  = {10},
  number  = {1},
  pages   = {21014},
  year    = {2020},
  doi     = {10.1038/s41598-020-77147-2}
}

@article{yu2024daseventnet,
  author  = {Yu, Pengliang and Zhu, Tieyuan and Marone, Chris and Elsworth, Derek and Yu, Mingzhao},
  title   = {{DASEventNet}: {AI}-based microseismic detection on distributed acoustic sensing data from the {Utah FORGE} well {16A} (78)-32 hydraulic stimulation},
  journal = {Journal of Geophysical Research: Solid Earth},
  volume  = {129},
  number  = {9},
  pages   = {e2024JB029102},
  year    = {2024},
  doi     = {10.1029/2024JB029102}
}

@article{tan2025aqueduct,
  author  = {Tan, Dao-Yuan and Tang, Zhen-Yu and Yan, Zhen-Rui and Wang, Jing and Zhang, Wei and Huang, Jing-Wu and Wang, Peng and Yuan, Zhiguo and Duan, Huan-Feng and Shi, Bin and Zhu, Hong-Hu},
  title   = {Real-time monitoring of water states in large-diameter aqueducts -- learning from distributed acoustic sensing signals},
  journal = {Communications Engineering},
  volume  = {4},
  number  = {1},
  pages   = {156},
  year    = {2025},
  doi     = {10.1038/s44172-025-00483-6}
}

@inproceedings{sandler2018mobilenetv2,
  author    = {Sandler, Mark and Howard, Andrew and Zhu, Menglong and Zhmoginov, Andrey and Chen, Liang-Chieh},
  title     = {{MobileNetV2}: Inverted Residuals and Linear Bottlenecks},
  booktitle = {Proceedings of the IEEE Conference on Computer Vision and Pattern Recognition ({CVPR})},
  pages     = {4510--4520},
  year      = {2018}
}

@inproceedings{huang2017densenet,
  author    = {Huang, Gao and Liu, Zhuang and van der Maaten, Laurens and Weinberger, Kilian Q.},
  title     = {Densely Connected Convolutional Networks},
  booktitle = {Proceedings of the IEEE Conference on Computer Vision and Pattern Recognition ({CVPR})},
  pages     = {4700--4708},
  year      = {2017}
}

@inproceedings{he2016resnet,
  author    = {He, Kaiming and Zhang, Xiangyu and Ren, Shaoqing and Sun, Jian},
  title     = {Deep Residual Learning for Image Recognition},
  booktitle = {Proceedings of the IEEE Conference on Computer Vision and Pattern Recognition ({CVPR})},
  pages     = {770--778},
  year      = {2016}
}

@inproceedings{ronneberger2015unet,
  author    = {Ronneberger, Olaf and Fischer, Philipp and Brox, Thomas},
  title     = {{U-Net}: Convolutional Networks for Biomedical Image Segmentation},
  booktitle = {Medical Image Computing and Computer-Assisted Intervention -- {MICCAI} 2015},
  series    = {Lecture Notes in Computer Science},
  volume    = {9351},
  pages     = {234--241},
  publisher = {Springer},
  year      = {2015},
  doi       = {10.1007/978-3-319-24574-4_28}
}

@article{munoz2022nearfield,
  author  = {Mu\~noz, Felipe and Soto, Marcelo A.},
  title   = {Enhancing fibre-optic distributed acoustic sensing capabilities with blind near-field array signal processing},
  journal = {Nature Communications},
  volume  = {13},
  pages   = {4019},
  year    = {2022},
  doi     = {10.1038/s41467-022-31681-x}
}

@article{lapins2024dasn2n,
  author  = {Lapins, Sacha and Butcher, Antony and Kendall, J.-Michael and Hudson, Thomas S. and Stork, Anna L. and Werner, Maximilian J. and Gunning, Jemma and Brisbourne, Alex M.},
  title   = {{DAS-N2N}: machine learning distributed acoustic sensing ({DAS}) signal denoising without clean data},
  journal = {Geophysical Journal International},
  volume  = {236},
  number  = {2},
  pages   = {1026--1041},
  year    = {2024},
  doi     = {10.1093/gji/ggad460}
}

@article{jousset2022volcanic,
  author  = {Jousset, Philippe and Currenti, Gilda and Schwarz, Benjamin and others},
  title   = {Fibre optic distributed acoustic sensing of volcanic events},
  journal = {Nature Communications},
  volume  = {13},
  pages   = {1753},
  year    = {2022},
  doi     = {10.1038/s41467-022-29184-w}
}

@article{shi2021transfer,
  author  = {Shi, Yi and Li, Yinghuan and Zhang, Yingchao and Zhuang, Zhemin and Jiang, Tao},
  title   = {An Easy Access Method for Event Recognition of {$\Phi$-OTDR} Sensing System Based on Transfer Learning},
  journal = {Journal of Lightwave Technology},
  volume  = {39},
  number  = {13},
  pages   = {4548--4555},
  year    = {2021},
  doi     = {10.1109/JLT.2021.3070583}
}

@article{zhang2024adai,
  author  = {Zhang, Shixiong and Li, Hao and Fan, Cunzheng and Zeng, Zhichao and Xiong, Chao and Wu, Jie and Yan, Zhijun and Liu, Deming and Sun, Qizhen},
  title   = {Adaptive decentralized {AI} scheme for signal recognition of distributed sensor systems},
  journal = {Opto-Electronic Advances},
  volume  = {7},
  pages   = {240119},
  year    = {2024},
  doi     = {10.29026/oea.2024.240119}
}

@article{sawi2026magnitude,
  author  = {Sawi, T. M. and McGuire, J. J. and Barbour, A. J. and Yoon, C. E. and Karrenbach, M. and Stewart, C.},
  title   = {Rapid earthquake magnitude classification via {P}-wave strains from borehole strainmeters and Distributed Acoustic Sensing},
  journal = {Nature Communications},
  volume  = {17},
  number  = {1},
  pages   = {4776},
  year    = {2026},
  doi     = {10.1038/s41467-026-72223-z}
}

@article{walter2020glacier,
  author  = {Walter, Fabian and Gr{\"a}ff, Dominik and Lindner, Florian and Paitz, Peter and K{\"o}pfli, Manuela and Chmiel, Matthias and others},
  title   = {Distributed acoustic sensing of microseismic sources and wave propagation in glaciated terrain},
  journal = {Nature Communications},
  volume  = {11},
  pages   = {2436},
  year    = {2020},
  doi     = {10.1038/s41467-020-15824-6}
}

@article{lin2024oceancurrents,
  author  = {Lin, J. and Fang, S. and He, R. and Tang, Q. and Qu, F. and Wang, B. and Xu, W.},
  title   = {Monitoring ocean currents during the passage of {Typhoon Muifa} using optical-fiber distributed acoustic sensing},
  journal = {Nature Communications},
  volume  = {15},
  pages   = {1111},
  year    = {2024},
  doi     = {10.1038/s41467-024-45412-x}
}

@article{li2023focalmechanisms,
  author  = {Li, Jiaxuan and Zhu, Weiqiang and Biondi, Ettore and others},
  title   = {Earthquake focal mechanisms with distributed acoustic sensing},
  journal = {Nature Communications},
  volume  = {14},
  pages   = {4181},
  year    = {2023},
  doi     = {10.1038/s41467-023-39639-3}
}

@article{tang2026intrusion,
  author  = {Tang, Rong and Rao, Yunjiang and Wu, Huijuan and others},
  title   = {Seamless integration of distributed acoustic sensing and passive optical networks for human intrusion monitoring},
  journal = {Communications Engineering},
  volume  = {5},
  pages   = {105},
  year    = {2026},
  doi     = {10.1038/s44172-026-00662-z}
}

@article{hernandez2022dasearthquake,
  author  = {Hern{\'a}ndez, Pablo D. and Ram{\'\i}rez, Jaime A. and Soto, Marcelo A.},
  title   = {Deep-learning-based earthquake detection for fiber-optic distributed acoustic sensing},
  journal = {Journal of Lightwave Technology},
  volume  = {40},
  number  = {8},
  pages   = {2639--2650},
  year    = {2022},
  doi     = {10.1109/JLT.2021.3138724}
}

@article{roumpos2024silicon,
  author  = {Roumpos, Ioannis and De Marinis, Lorenzo and Kovaios, Stefanos and Kincaid, Peter Seigo and Paolini, Emilio and Tsakyridis, Apostolos and Moralis-Pegios, Miltiadis and Berciano, Mathias and Ferraro, Filippo and Bode, Dieter and Srinivasan, Srinivasan Ashwyn and Pantouvaki, Marianna and Andriolli, Nicola and Contestabile, Giampiero and Pleros, Nikos and Vyrsokinos, Konstantinos},
  title   = {Silicon integrated photonic-electronic neuron for noise-resilient deep learning},
  journal = {Optics Express},
  volume  = {32},
  number  = {20},
  pages   = {34264--34274},
  year    = {2024},
  doi     = {10.1364/OE.532306}
}

@article{chen2017eyeriss,
  author  = {Chen, Yu-Hsin and Krishna, Tushar and Emer, Joel S. and Sze, Vivienne},
  title   = {{Eyeriss}: An Energy-Efficient Reconfigurable Accelerator for Deep Convolutional Neural Networks},
  journal = {IEEE Journal of Solid-State Circuits},
  volume  = {52},
  number  = {1},
  pages   = {127--138},
  year    = {2017},
  doi     = {10.1109/JSSC.2016.2616357}
}

@article{jin2024silicon,
  author  = {Jin, Zhicheng and Chen, Jiageng and Chang, Yanming and Liu, Qingwen and He, Zuyuan},
  title   = {Silicon photonic integrated interrogator for fiber-optic distributed acoustic sensing},
  journal = {Photonics Research},
  volume  = {12},
  number  = {3},
  pages   = {465--473},
  year    = {2024},
  doi     = {10.1364/PRJ.512298}
}

@article{zhang2026robustmicrophone,
  author  = {Zhang, Xuzhe and Tao, Yu and Luo, Xueqin and Wan, Yangyang and Yu, Xuhui and He, Zuyuan},
  title   = {A Mechanically Robust and Highly Sensitive Fiber-Optic Microphone},
  journal = {IEEE Sensors Journal},
  volume  = {26},
  number  = {14},
  pages   = {20973--20982},
  year    = {2026},
  doi     = {10.1109/JSEN.2026.3702683}
}

@misc{bengio2013estimating,
  author        = {Bengio, Yoshua and L{\'e}onard, Nicholas and Courville, Aaron},
  title         = {Estimating or Propagating Gradients Through Stochastic Neurons for Conditional Computation},
  year          = {2013},
  howpublished  = {Preprint at \url{https://arxiv.org/abs/1308.3432}}
}

@inproceedings{theis2017lossy,
  author    = {Theis, Lucas and Shi, Wenzhe and Cunningham, Andrew and Husz{\'a}r, Ferenc},
  title     = {Lossy Image Compression with Compressive Autoencoders},
  booktitle = {International Conference on Learning Representations},
  year      = {2017},
  url       = {https://arxiv.org/abs/1703.00395}
}

@inproceedings{krull2019noise2void,
  author    = {Krull, Alexander and Buchholz, Tim-Oliver and Jug, Florian},
  title     = {{Noise2Void}: Learning Denoising From Single Noisy Images},
  booktitle = {Proceedings of the IEEE/CVF Conference on Computer Vision and Pattern Recognition},
  year      = {2019},
  doi       = {10.1109/CVPR.2019.00223}
}

\end{document}